\documentclass[a4paper]{article}

\usepackage{INTERSPEECH2020}

\title{Exploration of End-to-end Synthesisers for \\ Zero Resource Speech Challenge 2020}
\name{Karthik Pandia D S, Anusha Prakash, Mano Ranjith Kumar M, Hema A Murthy}
\address{
  Indian Institute of Techonology Madras, India}
\email{pandia@cse.iitm.ac.in, anushaprakash@smail.iitm.ac.in, mano1381997@gmail.com, hema@cse.iitm.ac.in}

\begin{document}

\maketitle
\begin{abstract}

A Spoken dialogue system for an unseen language is referred to as Zero resource speech. It is especially beneficial for developing applications for languages that have low digital resources. Zero resource speech synthesis is the task of building text-to-speech (TTS) models in the absence of transcriptions. 

In this work, speech is modelled as a sequence of transient and steady-state acoustic units, and a unique set of acoustic units is discovered by iterative training. Using the acoustic unit sequence, TTS models are trained. 

The main goal of this work is to improve the synthesis quality of zero resource TTS system. Four different systems are proposed. All the systems consist of three stages-- unit discovery, followed by unit sequence to spectrogram mapping, and finally spectrogram to speech inversion. Modifications are proposed to the spectrogram mapping stage. These modifications include training the mapping on voice data, using x-vectors to improve the mapping, two-stage learning, and gender-specific modelling. Evaluation of the proposed systems in the Zerospeech 2020 challenge shows that quite good quality synthesis can be achieved.

\end{abstract}

\section{Introduction}

An infant learns the acoustic units of a language and reproduces them by babbling, even before s/he starts to recognise the sounds. The goal of Zerospeech task is to design a system that closely mimics this process. The Zerospeech challenge ultimately aims to build an autonomous speaker dialogue system. Previous challenges have focused on tasks such as spoken term detection, spoken term discovery (STD), and zero resource speech synthesis. The Zerospeech 2020 challenge consolidates the STD and synthesis tasks. The two tasks are run as task1 and task2, respectively. Acoustic unit discovery (AUD) is a common step in both tasks. STD and speech synthesis are applications of the discovered acoustic units (AUs).

Several AUD approaches have been proposed in the literature for various tasks \cite{pandia2019}.  The baseline provided by the organisers is Dirichlet process Gaussian mixture models (DPGMM) \cite{ondel2016variational} for AUD and Ossian \cite{wu2016merlin} for speech synthesis. 
Some of the classic AUD approaches are non-parametric Bayesian approach by Lee et. al. \cite{lee2012nonparametric}, AUD based on weak top-down constraints \cite{jansen2013weak}, and Autoencoder-based approaches \cite{badino2014auto, renshaw2015comparison}. Most systems in the Zerospeech 2019 challenge \cite{Dunbar2019} used autoencoder-based approaches for AUD \cite{Eloff2019, Liu2019, Tjandra2019}. The objective function in most approaches is frame-based, wherein sequence information is not explicitly modelled. Hidden Markov model (HMM) based generative modelling of AU is one of the top systems \cite{pandia2019} in terms of the synthesis quality and low-bitrate encoding. Even the baseline is a non-parametric HMM-based approach \cite{ondel2016variational}. In \cite{pandia2019}, AUs are modelled explicitely as transient and steady-state regions. A modified version of this approach is used in this work. In the past two years, there has been a surge in the development of training approaches for TTS \cite{oord2016wavenet,tacotron2, waveglow, li2018close}. The current work uses the AUD technique from \cite{pandia2019} and focuses on improving the synthesis quality in an end-to-end framework.

The TTS framework used in this work has two stages. The first stage maps the AU symbol sequence to the corresponding spectrogram. The second stage inverts the spectrogram back to speech. Since the task is to synthesise speech in a target speaker's voice, the inversion stage is fixed. Different methods are proposed to learn the mapping between the AU sequence and spectrogram efficiently. The proposed approaches explore an end-to-end framework that includes speaker embedding, hierarchical training, and gender-dependent training to learn the mapping. 

End-to-end TTSes are trained based on the Tacotron2 architecture \cite{tacotron2}. x-vectors are used as speaker embedding to produce speech in the target speaker's voice \cite{espnet}. In hierarchical learning, similar to the AUD approach, the mapping is first confined to smaller units. The obtained model is then used to bootstrap the learning on full utterances. Since the training data has both male and female speakers, gender-dependent TTSes are developed. During synthesis, depending on the gender of the target speaker, the appropriate TTS is employed. Subjective measures indicate that there is an improvement in the overall quality of the synthesised speech output compared to our systems in Zerospeech 2019 challenge, with a slight degradation in the speaker similarity measure.

    The rest of the paper is organised as follows. The proposed approaches are presented in Section \ref{sec:proposed}. Section \ref{sec:experiments} details the experiments carried out. Section \ref{sec:results} discusses the results and their analysis. The work is concluded in Section \ref{sec:conclusion}.

\section{Proposed systems}
\label{sec:proposed}

The AUD approach used in this work is similar to the approach in ~\cite{pandia2019}. A modification to the syllable-like segmentation algorithm is proposed, which makes AUD totally unsupervised. The AUD approach and the four TTS systems incorporated in the zerospeech pipeline are explained in this section. These systems are illustrated by a block diagram given in Figure \ref{fig:systems_diagram}.


\subsection{Acoustic unit discovery (AUD)}

\begin{figure}[t]
  \centering
  \includegraphics[width=\linewidth]{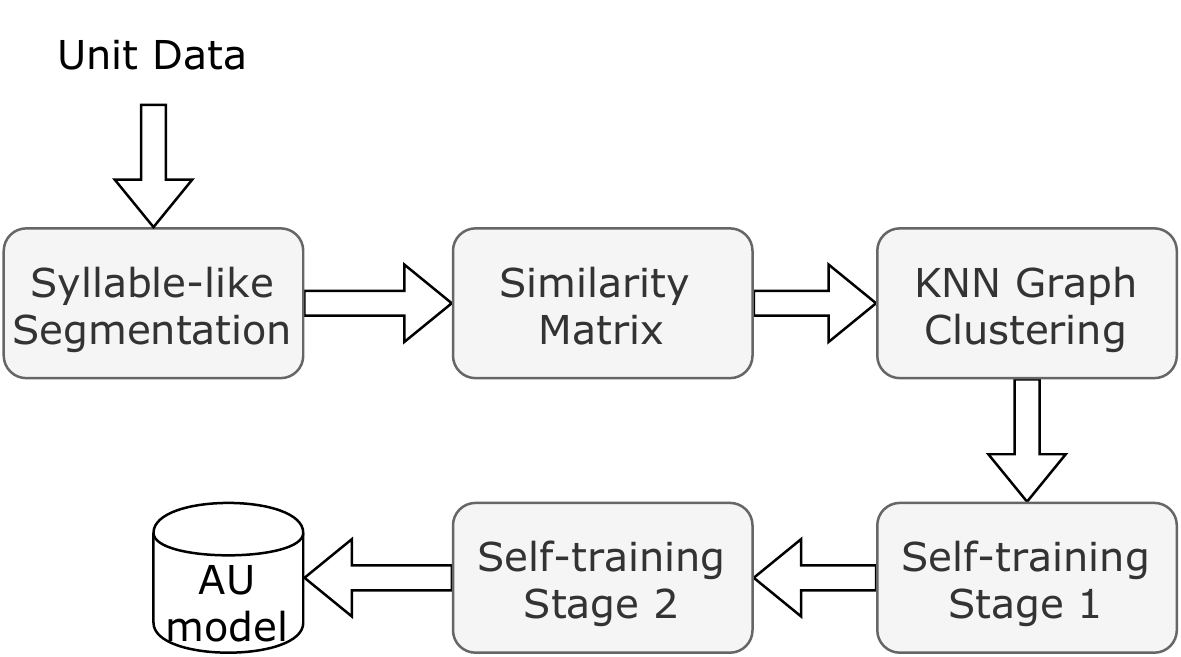}
  \caption{Steps for acoustic unit discovery (AUD)}
  \label{fig:AUD_diagram}
\end{figure}

An overview of the AUD approach used in ~\cite{pandia2019} is briefly given here. The approach models speech as transient and steady-state regions. The transient regions correspond to rising and falling transients, and the steady-state regions predominantly correspond to vowels.  The block diagram of the proposed AUD approach is shown in Figure~\ref{fig:AUD_diagram}.  First, speech is segmented into syllable-like units. A similarity matrix is obtained by computing the DTW score between all pairs of syllable-like segments. Homogeneous syllable like units are clustered using a K-nearest neighbour (KNN) graph clustering approach. 
The syllable-like units in each cluster is a sequence of 3 AUs corresponding to rising transient, steady-state, and falling transient. HMMs are used to model the AUs. Using the trained models, the syllable-like units are transcribed. Then the obtained transcriptions are used to retrain models. The training and transcription processes are repeated until convergence. This process of repeated train and transcribe is termed as \textit{self training}. The initial models thus obtained are trained only on the syllable-like segments present in the clusters. 
Using the initial models, the full set of syllable-like segments are transcribed. Self-training is performed on this set to obtain better models. The self-training process on the syllable-like segments is referred to as stage 1 training. 
Once the models are trained on the entire set of syllable-like segments, the models are used to transcribe continuous speech. Stage 2 self-training is performed on continuous speech until convergence. The models thus obtained are the final models used to generate the AU sequence.

In \cite{pandia2019}, a vowel posterior function was used to extract syllable-like units, which is not completely unsupervised. In the current work, a signal processing based approach is used to segment the speech into syllable-like regions, making it completely unsupervised. Specifically, the unsupervised segmentation approach uses short-time energy (STE) post-processed using a group-delay function \cite{prasad2004automatic}. The STE function is assumed as a magnitude function of a signal. It has been shown that the poles and zeros of a magnitude function can be better resolved using group-delay processing by deriving the minimum phase function corresponding to the original signal \cite{sebastian2016analysis}. Accordingly, a function is derived, and the group-delay is computed to resolve the peaks and valleys of the STE reliably.  

\subsection{Text-to-speech (TTS) synthesis (System 1)}

For synthesising speech, the end-to-end paradigm is used. The end-to-end speech synthesis framework is an attractive platform to use as training TTSes is easy. It alleviates the need for separate modules for feature engineering and language-specific tasks. Synthesisers can be trained given only speech waveforms and corresponding text transcriptions (sequence of acoustic units in this case). 

The end-to-end framework used in this work is based on the Tacotron2 architecture \cite{tacotron2}. It takes care of the conversion of a sequence of AUs to mel-spectrograms. Tacotron2 consists of an encoder and a decoder with attention weights. The encoder extracts sequential information from the character embeddings, and the attention module predicts a fixed-length context vector. The decoder predicts frame-level mel-spectrograms at each step.

\begin{figure}[t]
  \centering
  \includegraphics[width=1\linewidth]{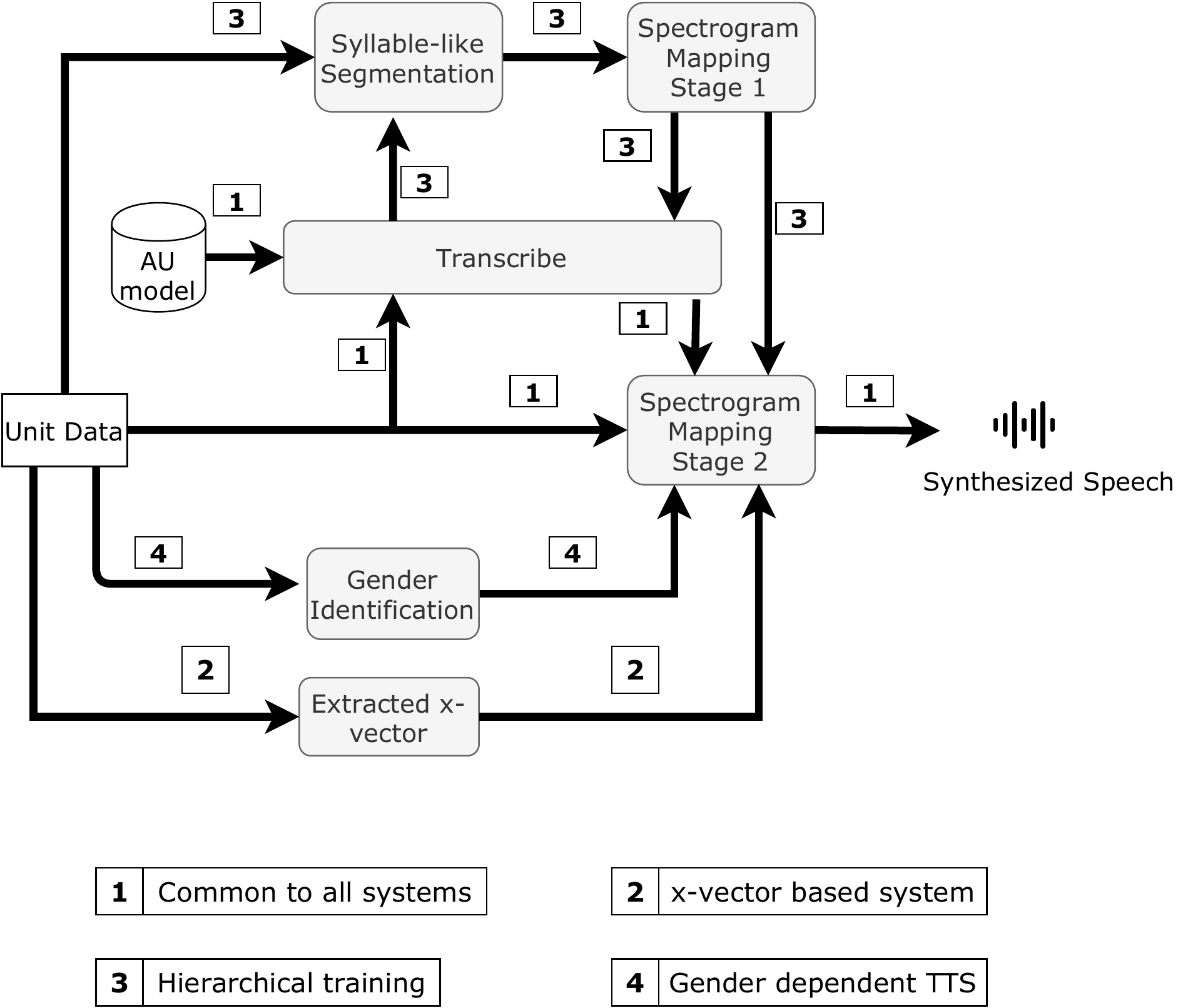}
  \caption{Block diagram of the proposed systems. Numbers next to the arrows refer to the flow of the corresponding system type. The flow of System type 1 is common to all other systems.}
  \label{fig:systems_diagram}
\end{figure}

For the speech waveform inversion, the WaveGlow vocoder is used \cite{waveglow}. WaveGlow takes in mel-spectrogram as input and generates the speech output. WaveGlow is a neural vocoder which is a combination of WaveNet \cite{oord2016wavenet} and Glow \cite{Glow_NIPS2018}. It uses a single network with a single likelihood cost function. WaveGlow is a generative model, in which samples are generated from a zero mean spherical Gaussian, whose dimension is the same as that of the output. It uses a series of non-linear layers to transform the Gaussian distribution to the desired distribution. The desired distribution comes from audio samples conditioned on mel-spectrograms. The flow of this system can be seen in Figure \ref{fig:systems_diagram}. It is to be noted that this flow is common to all the other proposed systems.

\subsection{TTS with speaker embedding (System 2)}

One of the objectives of the challenge is to produce synthesised speech in the target voice. The conventional Tacotron2 framework does not incorporate any speaker-specific information and may not be suited to a multi-speaker setting. Incorporating speaker embedding in the Tacotron2 framework provides better flexibility in terms of speaker selection. For this purpose, x-vectors are used. x-vectors have been conventionally used for speaker recognition and verification tasks \cite{xvector_ASR, xvector_SV2017}, and have now been applied to TTS tasks too \cite{espnet}.

x-vectors are fixed-length speaker embeddings computed from variable length utterances. The model to compute x vector is trained to discriminate speakers using a time-delay neural network (TDNN) architecture \cite{peddinti2015time}. x-vectors are extracted from the audio files and then appended to each encoder state of the sequence-to-sequence model. The TTS synthesiser is then trained \cite{espnet}.

During testing of the TTS, per utterance x-vector is not available as only transcriptions are provided. Hence, the mean of x-vectors corresponding to the speaker in the training data is considered as the speaker x-vector. The same speaker x-vector is appended to all the encoder states for synthesis.

\subsection{Hierarchical training (System 3)}

In a conventional Tacotron TTS system, the mapping between the symbols and spectrogram is learnt at the utterance level. It has been shown in the literature that when the training is performed on short utterances, the model learns the mapping better. This is primarily because the confusion in mapping between the input symbol sequence and the spectrogram decreases. Unlike a phoneme or character that is predefined for a language, there is an inherent confusion between the discovered AUs. This is because unlike phonemes, AUs are not uniquely represented for different sounds. This confusion adds to the existing problem of training with long utterances. To alleviate these issues, the mapping between the AU sequence and spectrogram regions is performed at the syllable level. This is similar to unit discovery training used in this work, wherein the initial self-training first starts at the syllable level. Similar to AUD, learning of the mapping is constrained within smaller syllable-like segments, leading to robust initial models. The initial network parameters thus obtained are used to bootstrap the training process in the next stage, where fine-tuning is performed using the utterance-level data.

\subsection{Gender-dependent TTS (System 4)}

One of the objectives of Zerospeech TTS task is to synthesise speech in a target speaker's voice. The characteristics of the target speaker have to be preserved while training. There are two components in the system - symbols to spectrogram mapping and spectrogram to audio inversion. Similar to the phoneme, the AU should be agnostic to the speaker's characteristics. Therefore, the target speaker's data can be used to train both the mapping and inversion task. Surprisingly, it is observed that the synthesised speech has the source speaker's characteristics. This means that the source speaker's characteristics in terms of timbre are embedded in the spectrogram, through the symbol sequence.

Experiments are conducted to evaluate the difference in the quality of the synthesised speech when (a) unit data, (b) voice data is used for spectrogram mapping. When the data provided for unit discovery was used to learn the spectrogram mapping, there was an improvement in the intelligibility of the synthesised speech. This improvement in intelligibility can be attributed to the amount of data used to learn the mapping. Although the intelligibility improved, the speaker similarity measure degraded. This was more pronounced when male speaker's audio was synthesised using female speaker's voice and vice-versa. Hence, gender-dependent systems are built, which take care of both data insufficiency and the problem of mismatched condition.

Gender identification is performed using GMMs. Mel frequency cepstral coefficients using longer frame size are extracted. A GMM is trained using both male and female target speakers' data. Then the male and female GMMs are trained using maximum aposteriori adaptation. During classification, the likelihood ratio (LR) testing is performed. Each file in the unit set is classified based on LR score. A voting rule is applied to this decision to arrive at a final decision of gender classification. The gender with a high number of votes is classified as the identified gender. Once the gender of the unit files is identified, spectrogram mapping is performed separately for each gender. During testing, the appropriate spectrogram mapping model is used to estimate the spectrogram, which is then fed to the WaveGlow model for synthesis.


\section{Experiments}
\label{sec:experiments}

\subsection{Dataset}

The datasets used in this work are part of the Zerospeech 2019 challenge data. Datasets are provided for two languages-- English, the development language, and Indonesian, the surprise language. The dataset for each language is divided into three sets. Unit set to train AU models, voice set to train TTS model, and test set to evaluate the system.
The English dataset has about 15 hours unit data (100 speakers); 2 hours (V001, male) and 2.6 hours (V002, female) of voice data for target speakers. The test data has about 28 minutes of data across 34 speakers.
The Indonesian dataset (surprise language) has about 15 hours of unit data (112 speakers); 1.5 hours of voice data for a target female speaker. The test data has about 29 minutes of data across 15 speakers. A detailed description of the surprise language dataset is given in \cite{sakti2008development1, sakti2008development2}. The task is to synthesise test sentences uttered by a source speaker with the target speaker's characteristics. This is similar to the voice conversion problem.

\subsection{Systems}

The four systems developed as part of this challenge differ in their synthesiser. The AUD approach is the same across all systems. Kaldi toolkit \cite{Povey_ASRU2011} is used for AUD. ESPnet, an implementation of Tacotron2, is used for AU sequence to mel-spectrogram conversion \cite{espnet}. The encoder-decoder network is trained for 200 epochs using location-sensitive attention along with guided attention. Two variations of System 1 are built-- System 1 (unit) using unit data, and System 1 (voice) using voice data.


WaveGlow uses mel-spectrograms extracted with 80 bins librosa \cite{librosa} mel filters. Training a good WaveGlow model from scratch is time-consuming. Hence, in this work, WaveGlow models are re-trained on the pre-trained LJ Speech model \cite{waveglow} for about 10K iterations. Three WaveGlow models, corresponding to target speakers, are trained on the voice data-- two for English (V001, V002) and one for the surprise language.




 For speaker embedding, 512-dimensional x-vectors are extracted from the audio files using a pre-trained x-vector \cite{xvector_ASR} provided by Kaldi. For this variety of TTS, even the voice train data is pooled along with the multi-speaker source data for training.

Test sentences were synthesised across all systems. Listening tests suggested that System 1 (voice) and System 2, corresponding to vanilla TTS trained on voice data and TTS with speaker embedding, respectively, were the best systems. Hence, Systems 1 (voice) and 2 were submitted to the Zerospeech 2020 Challenge.

 

\section{Results and analysis}
\label{sec:results}

\begin{table}[h]
\centering
\caption{Mushra test scores of different systems}
\label{tab:mushra}
\begin{tabular}{|c|c|}
\hline
   & \textbf{Score} \\ \hline
\textbf{System 1 (unit)} &  46.77         \\ \hline
\textbf{System 1 (voice)} &  43.13          \\ \hline
\textbf{System 2} &  44.29          \\ \hline
\textbf{System 4} &  33.67       \\ \hline
\end{tabular}
\end{table}

\begin{table*}[h]
\caption{ Evaluation measures on development and test languages. Scores are shown for the baseline system, topline system, the two systems submitted to Zerospeech 2019 (ZS19) and Zerospeech 2020 (ZS20) challenges}
\label{tab:table1}
\begin{tabular}{|l|c|c|c|c|c|c|c|c|c|c|}
\hline
\multicolumn{1}{|c|}{\textbf{}}        & \multicolumn{5}{c|}{\textbf{Development Language (English)}}                        & \multicolumn{5}{c|}{\textbf{Test Language (Indonesian)}}                         \\ \hline
\multicolumn{1}{|c|}{\textbf{Systems}} & \textbf{MOS} & \textbf{CER} & \textbf{Similarity} & \textbf{ABX} & \textbf{Bitrate} & \textbf{MOS} & \textbf{CER} & \textbf{Similarity} & \textbf{ABX} & \textbf{Bitrate} \\ \hline
\textbf{Baseline}                      & 2.14         & 0.77         & 2.98                & 35.63        & 71.98            & 2.23         & 0.67         & 3.26                & 27.46        & 74.55            \\ 
\textbf{Topline}                      & 2.52       & 0.43         & 3.1                & 29.85        & 37.73            & 3.49         & 0.33         & 3.77                & 16.09        & 35.2            \\ \hline
\textbf{System 1 (voice) (ZS20)}                      & 3.15         & 0.61         & 2.69                & 33.28        & 126.41           & 3.09         & 0.67         & 2.45                & 34.33        & 101              \\ 
\textbf{System 2 (ZS20)}                      & 3.19         & 0.45         & 2.68                & 33.28        & 126.41           & 3.66         & 0.44         & 2.49                & 34.33        & 101              \\ \hline
\textbf{System 1 (ZS19)}      & 2.82         & 0.55         & 2.76                & 29.66        & 138.59           & 2.53         & 0.43         & 3.58                & 23.56        & 115.43           \\ 
\textbf{System 2 (ZS19)}      & 2.77         & 0.61         & 3                   & 28.16        & 92.75            & 2.02         & 0.48         & 3.21                & 20.77        & 94.15            \\ \hline
\end{tabular}
\end{table*}

Initially, an evaluation is conducted to assess the performance of different unit sequence-to-spectrogram mapping approaches. It is carried out only for English dataset. Mushra test \cite{recommendation2001method} is performed using $10$ random utterances from the test set. The target utterance is provided as reference audio, and the ratings for the system outputs for different systems are given on a scale of 0 to 100, 100 being the best. 18 listeners participated in the evaluations. The results are shown in Table~\ref{tab:mushra}.

For System 3 (hierarchical), it was observed that some of the synthesised utterances had artefacts and random speech in certain regions. Although the syllable-level mapping was learnt well, this didn't scale to the utterance-level. The reason for such artefacts needs to be further investigated. Hence, it was not considered for Mushra evaluation.

System 1 (unit) gave the best Mushra score of $46.77$, followed by System 2 (x-vector), and System 1 (voice). Although System 2 and System 1 (voice) had better speaker similarity, the overall intelligibility of System 1 (unit) was better. Hence, the listeners ignored the speaker factor while giving the scores. The performance of System 4 was poor as synthesised audio had artefacts, though not to the extent present in the synthesised audio of System 3. Based on the Mushra test scores in development language (English) and based on inspection of speaker similarity, System 1 (voice) and System 2 were submitted for the challenge.


A detailed analysis of the results of the systems submitted to Zerospeech 2019 (ZS19) and 2020 (ZS20) challenges is presented here. In our submission to ZS19, the focus of the experiments was on unit discovery. In the current work, experiments are extensively conducted to improve the synthesis quality by fixing the AUD method. In simple words, the sole objective is to improve the subjective evaluations: mean opinion score (MOS) and character error rate (CER).

Table~\ref{tab:table1} summarises the results of the baseline and topline systems of ZS20, and the results of our submissions to ZS19 and ZS20. First, we compare results of ZS20 systems-- System 1 (voice) and System 2.
According to Table~\ref{tab:mushra}, the overall synthesis quality of System 2 is better than System 1 (voice) by $(1.16/100)$. The scores (Table~\ref{tab:table1}: MOS) of the systems on development data also show a similar pattern. But for Indonesian, the difference is significant $(0.57/3)$. 
The improvement in CER is significant when the x-vector model (System 2) is used for spectrogram mapping. The absolute improvement of the measure is $16\%$ for English and $23\%$ for Indonesian. Hence, System 2 is better in terms of subjective evaluation measures, with more or less similar speaker similarity numbers.


In our ZS19 submission, it was observed that the MOS and CER scores were high when the number of AUs were larger, albeit with increased bitrate. This increase in bit rate was approximately 20 for an increase in the number of units from 40 to 112. For this year's work, the number of AUs is set to 100 so that the bit rate is still not large. Though the segmentation approach is different in this year's work, the bitrate follows a similar pattern. The bitrate is 126.41 for English and 101.0 for the Surprise language. 

When compared to last year's results (ZS19), the speaker similarities for all the current submissions are lower. This effect is seen even in the x-vector system (ZS20-System 2), which is supposed to normalize the speaker's characteristics with the help of speaker embeddings. This could mean that there is an inherent problem with spectrogram mapping, which is trained on unit data. Moreover, the source speaker's characteristics seem to exist even when spectrogram mapping is trained only on voice data. This problem might be because of data insufficiency for voice data, as E2E systems require a large amount of training data.

Comparing MOS results with the systems submitted in ZS19, there is an absolute increase of $0.4$ for English and $1.13$ for Indonesian. The CER also improves by $0.1$ for English, whereas the value remains almost the same for the Indonesian language. The effects on both languages do not seem to follow similar patterns. 

Based on the discussion and analysis of results, we see that the overall synthesis quality has improved compared to our submission to Zerospeech 2019 challenge, where the baseline TTS (Ossian) was used.
To get the full benefit of using E2E approaches, the proposed techniques have to be applied to a much larger training data.








\section{Conclusion}
\label{sec:conclusion}

While AUD is important towards building zero resource speech synthesis systems, the TTS component is equally vital. In this work, the AU sequence in terms of steady-state and transient regions in speech is used as transcription to build TTS systems. Various approaches to improve the synthesis quality in an end-to-end framework are explored. While there is a significant improvement in the overall synthesis quality using the E2E framework, speaker similarity seems to be an issue. Despite using only the target speaker data to train the TTS model, the source speaker's characteristics are observed in the synthesised speech. This problem could be addressed by increasing the amount of data for the target speaker.

\bibliographystyle{IEEEtran}

\bibliography{paper}

\end{document}